\documentclass[12pt,preprint]{aastex}

\shorttitle{}
\shortauthors{de Mello et al.}

\begin{document}


\title{Searching for Star Formation Outside Galaxies: Multiwavelength Analysis of the Intragroup Medium of HCG100}


\author{D. F. de Mello\altaffilmark{1,2,3} }

\author{S. Torres-Flores\altaffilmark{4}, \& C. Mendes de Oliveira\altaffilmark{4}}


\altaffiltext{1}{Observational Cosmology Laboratory, Code 665, Goddard Space Flight Center, Greenbelt, MD
20771}
\altaffiltext{2}{Catholic University of America Washington, DC 20064}

\altaffiltext{3}{Johns Hopkins University, Baltimore, MD 21218}

\altaffiltext{4}{Universidade de S\~ao Paulo, Instituto de Astronomia, Geof\' isica e Ci\^encias Atmosf\'ericas, Departamento de
astronomia, S\~ao Paulo, Brazil}


\begin{abstract}

We used multiwavelength data (H{\sc i}, FUV, NUV, R) to search for evidence 
of star formation in the intragroup medium of the Hickson Compact Group 100. 
We find that young star-forming regions are located in the intergalactic H{\sc i} clouds of the
compact group which extend to over 130 kpc away from the main galaxies. A tidal dwarf galaxy candidate
is located in the densest region of the H{\sc i} tail, 61 kpc from the brightest group member and its age is
estimated to be only 3.3 Myr. Fifteen other intragroup H{\sc ii} regions and TDG candidates are detected in the GALEX FUV image and within a field 
10$'$$\times$10$'$ encompassing the H{\sc i} tail. They have ages $<$200 Myr, 
H{\sc i} masses of 10$^{9.2-10.4}$ M$_{\odot}$, 
0.001$<$ SFR $<$0.01 M$_{\odot}$ yr$^{-1}$, and stellar masses 10$^{4.3}$--10$^{6.5}$ M$_{\odot}$. The H{\sc i} clouds to
which many of them are associated have column densities about one order
of magnitude lower than  N(H{\sc i})$\sim$10$^{21}$ cm$^{-2}$.

\end{abstract}


\keywords{groups of galaxies: general --- compact groups of galaxies: individual(HCG100)}



\section{Introduction}

The environment of galaxies play an important role on determining their overall properties. 
One of the most interesting environmental effects is seen in interacting systems which contain 
stripped H{\sc i} gas in the intragroup/intergalactic medium, instead of around galaxies. 
Three independent studies, Mendes de Oliveira et al. (2004, for the
Stephan's Quintet), Mendes de Oliveira et al. (2006, for HCG 31), Ryan-Weber et al. (2004, for NGC 1533, HCG 16, ESO
149-G003) and Oosterloo et al. (2004, for the loose group around NGC 1490) have surveyed systems 
with H{\sc i} intergalactic clouds and 
have shown that these are associated with actively star forming
regions, the so-called {\it intergalactic H{\sc ii} regions}. These objects seem to
be similar to the H{\sc ii} regions in our Milky Way, but are located
in the intragroup medium and have high metallicities (close
to solar in most cases). The fate of these types of objects and their importance in galaxy 
formation and evolution, enrichment of the intergalactic medium and
globular cluster formation is still debatable. In order to better address these issues, we 
embarked on a multiwavelength analysis of the environment of interacting systems and present here
the results based on FUV, NUV, optical, and H{\sc i} data of a Hickson compact group, HCG100.

The last group in Hickson's catalog of compact groups of galaxies (Hickson 1982) is at v$_{\rm R}$ = 5336 km/s (z=0.0178, Hickson et al. 1992) and
it is formed by four late-type galaxies with accordant redshifts: a bright central 
Sb galaxy (HCG100a), an irregular galaxy with an optical tidal tail (HCG100b), a late-type barred spiral (HCG100c) 
with central H$\alpha$ emission (Vilchez \& Iglesias-P\'aramo 1998)
and a late-type edge-on spiral (HCG100d). HCG100a, b and c show strong evidence of interaction, demonstrated 
not only by peculiarities in their morphologies but also in their velocity fields (Plana et al. 2003). 
H100b shows a tidal tail connecting with a faint galaxy not originally classified as a member of the group by
Hickson. Past encounters are also confirmed by an 
extended H{\sc i} tail (Verdes-Montenegro et al. 2006).
Therefore, the well-advanced dynamical state of HCG100 makes it an ideal target for searching for intergalactic H{\sc ii} regions that
might have been triggered by tidal interaction. 

This paper is organized as follows: \S 2 describes the data, \S 3 discusses the age, masses and star formation rate estimates, 
\S 3.1 and \S 3.2 compare the intragroup regions in HCG100 with other intergalactic H{\sc ii} regions and high$-z$ UV-luminous galaxies.
Finally, \S4  summarizes the main conclusions. Throughout this paper, we use a cosmology
with $\Omega_{\rm M}=0.3$, $\Omega_{\Lambda}=0.7$~and $h=0.7$.
Magnitudes are given in the AB-system and the adopted distance to HCG100 is 76.3 Mpc.

\section {The data}

\begin{figure}
\plotone{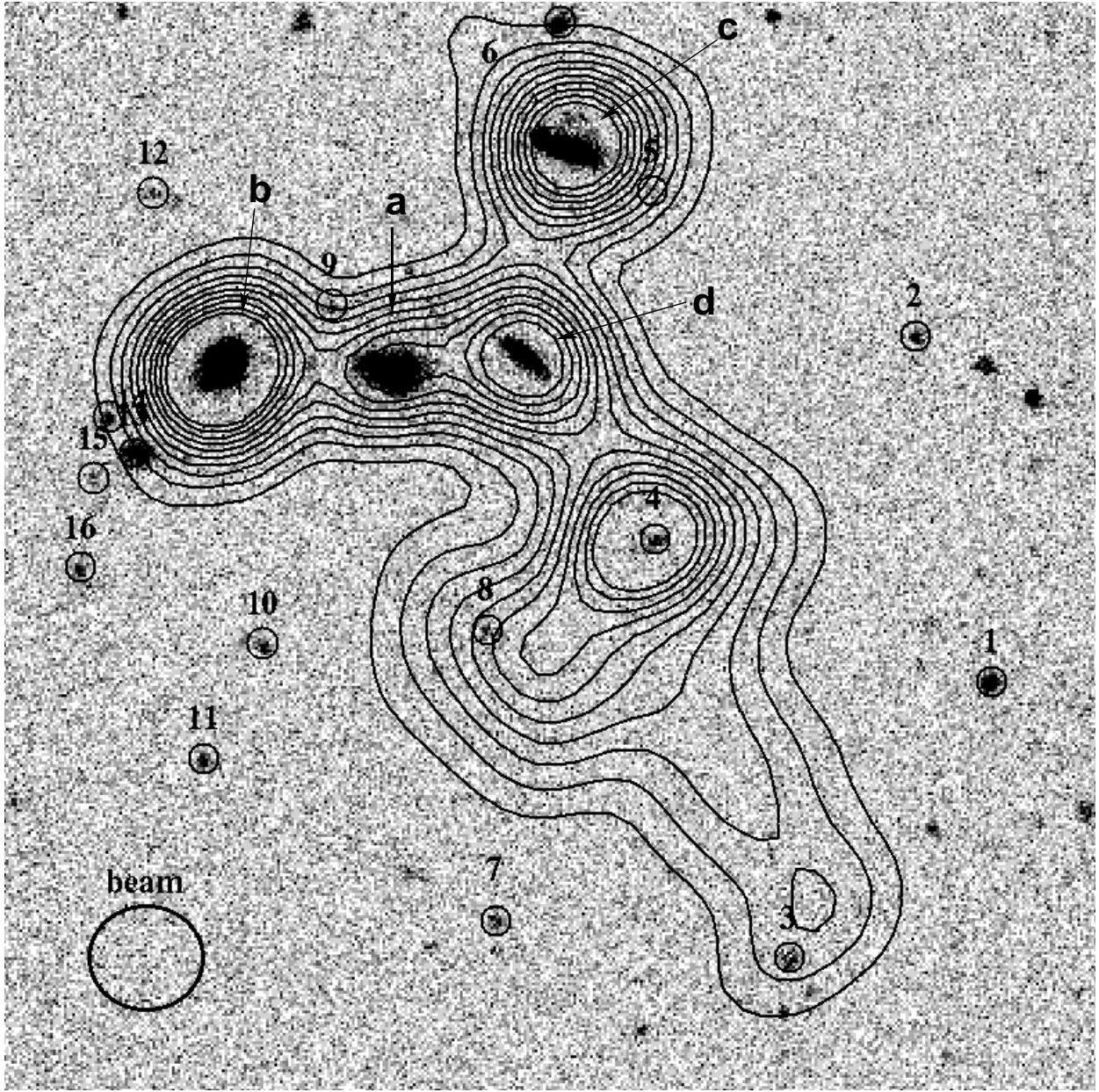}
\caption{GALEX FUV image 10$'$$\times$10$'$ with H{\sc i} contours. Four HCG100 members are marked as `a, b, c and d'. Intragroup objects are circled (radius=8$''$) and numbered. VLA NH{\sc i} contours are 
0.6, 1.2, 2.1, 3.6, 4.4, 5.1, 5.9, 6.6, 7.4 $\times$10$^{20}$ cm$^{-2}$, beam size (61.0$''$$\times$55.23$''$) is shown on left corner. North is
to the top and East to the left. 
\label{fuvhi_10arcmin}}
\end{figure}

\begin{figure}
\plotone{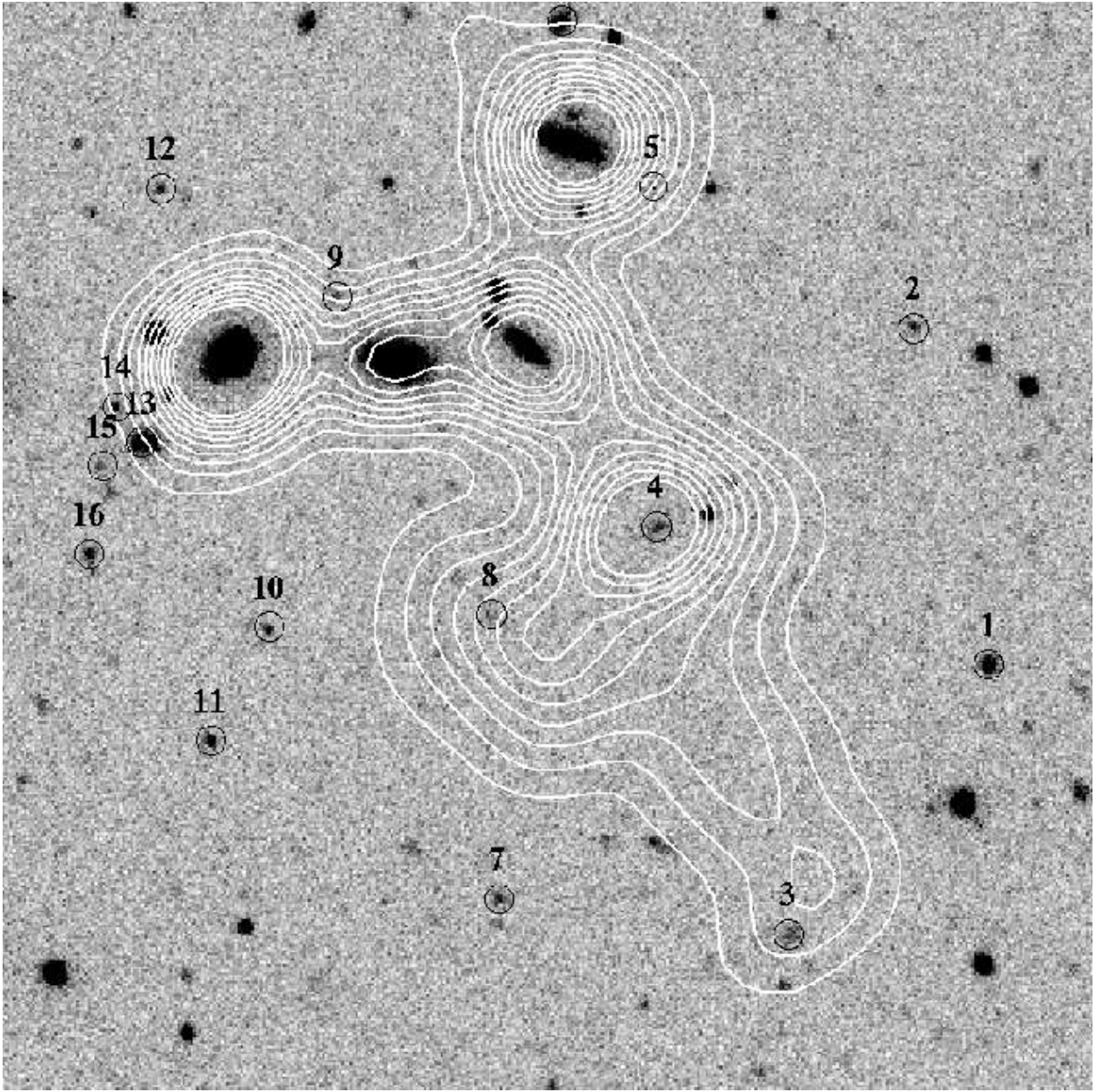}
\caption{GALEX NUV image 10$'$$\times$10$'$ with H{\sc i} contours. Intragroup objects are circled (radius=8$''$) and numbered. VLA NH{\sc i} contours are 
0.6, 1.2, 2.1, 3.6, 4.4, 5.1, 5.9, 6.6, 7.4 $\times$10$^{20}$ cm$^{-2}$. Beam size (61.0$''$$\times$55.23$''$) is
shown in Fig.\ref{fuvhi_10arcmin}. North is
to the top and East to the left.
\label{nuvhi_10arcmin}}
\end{figure}

HCG100 was observed with the Galaxy Evolution Explorer (GALEX) mission in the far and near ultraviolet (FUV $\lambda$$_{\rm eff}$=1528\AA, NUV $\lambda$$_{\rm eff}$=2271\AA) 
as part of our Cycle 1 program (GI\#73) to observe compact groups of galaxies. In Fig. \ref{fuvhi_10arcmin} and Fig.\ref{nuvhi_10arcmin} we show a 
cutout of a 10$'$$\times$ 10$'$ window of the FUV and NUV images 
(GALEX field of view is 1$^{\circ}$.28 and 1$^{\circ}$.24 in FUV and NUV, respectively and pixel 
scale is 1.5 arcsec pixel$^{-1}$) together with the distribution of atomic hydrogen H{\sc i} (PI Verdes-Montenegro).

We have also obtained an R-band image with the CTIO Blanco 4m telescope and a mosaic II CCD imager. Three 300s images of the
group were taken, at a median seeing of 1.1$''$.  Each frame covered a 40
arcmin field on a side, at a pixel scale of 0.27$''$/pixel, but we only used a 10
arcmin field around the object, given that this was the area of interest.
The data was bias subtracted and flat fielded using standard procedures
(using the package mscred in IRAF\footnote{IRAF is distributed by the National Optical Astronomy Observatories,
    which are operated by the Association of Universities for Research
    in Astronomy, Inc., under cooperative agreement with the National
    Science Foundation.}).  A flatfield was constructed from
a combination of dark sky frames and twilight flats which worked well,
given that the background, after flatfielding was done, showed an rms
variation no greater than 1\% over the whole field. The night was not
photometric and therefore no calibration stars were observed.  Instead, we
chose to use two galaxies previously observed by Rubin et al.  (1991),
HCG 100c and HCG 100d, for obtaining the zero point of the photometry
(see below).

\subsection{Flux Calibration}

FUV and NUV fluxes were calculated using Morrissey et al. (2005) m$_{\lambda}$=-2.5 log[F$_{\lambda}$/a$_{\lambda}$] + b$_{\lambda}$, 
where a$_{FUV}$ = 1.4 $\times$ 10$^{-15}$ erg s$^{-1}$ cm$^{-2}$ \AA$^{-1}$,  
a$_{NUV}$=2.06$\times$ 10$^{-16}$ erg s$^{-1}$ cm$^{-2}$ \AA$^{-1}$, b$_{FUV}$=18.82 and b$_{NUV}$=20.08 for FUV and NUV, respectively. Fluxes were multiplied
by the effective filter bandpass ($\Delta$$\lambda$$_{FUV}$=269\AA\, and $\Delta$$\lambda$$_{NUV}$=616\AA) to give units of erg s$^{-1}$ cm$^{-2}$ and  
luminosities were calculated for a distance D=76.3 Mpc.  

The R-band image was calibrated to match Rubin et al. (1991) magnitudes for galaxies HCG100c and HCG100d. Total integrated R magnitudes of these
objects are given in their Table 3 (no errors were quoted), and these were compared with the total
instrumental magnitudes we obtained for these galaxies using two methods:
(1) through aperture photometry in IRAF (using the task daophot.phot)
and (2) using the program SExtractor (Bertin and Arnouts et al. 1996).
The zero points obtained for each of these objects, and for each method,
agreed within 0.1 mag, and we used an average of the values as our final
zero point. The lack of a proper calibration, does not significantly change our results as shown in \S~3.

The H{\sc i} data taken with the VLA (configuration D, PI Verdes-Montenegro) is not able to resolve the intragroup objects 
since the beamwidth is 61.0$''$$\times$55.23$''$ (22.6~kpc $\times$ 20.4~kpc at 76.3~Mpc) and objects have sizes $<$16$''$. Therefore, our
measurements of the H{\sc i} fluxes were centered on the FUV detections but measured 
within one beam. Fluxes were converted into mass by using the relation:

\begin{equation} 
M_{HI}=2.356 \times 10^5 F_{HI} D^2
\end{equation}

where $D$ is the distance to the group in Mpc, $F_{HI}$ is the H{\sc i} flux in Jy km/s.

Column Densities were calculated by applying the following relation
\begin{equation} 
N_{HI}=1.82\times 10^{18} (\lambda ^2 / (2.65 \Theta ^2) F
\end{equation}

where $\Theta$=beamwidth in arcmin$^{2}$, $\lambda$ is the wavelength (21cm) and F is the H{\sc i} flux density in Jy/beam km/s 
(Rohlfs \& Wilson 2000).

\subsection{Source Detection}

We used SExtractor (SE, Bertin \& Arnouts 1996) to detect sources in the FUV and matched that catalog with the
NUV and R-band catalogs. Therefore, only objects with FUV detections were included in our final catalog. 
In this paper we concentrate on all objects which were detected in a region of 10$'$$\times$ 10$'$ centered on the H{\sc i} 
tail (Verdes-Montenegro et al. 2006). 

We used SE's Kron elliptical apertures to measure magnitudes (mag$_{-}$auto) in FUV, NUV and R-bands. Although  
mag$_{-}$auto is often used to measure total magnitudes in galaxy surveys (e.g. Bell et al. 2004,  de Mello et al. 2006, Zucca et al. 2006), 
high uncertainties might be expected at the faint magnitudes due to the assumption that the sky background has Gaussian random 
noise without source confusion (Brown et al. 2007). However, since we are comparing data taken with different 
resolutions and the fact that UV light does not necessarily peak at the same coordinate as the optical light, mag$_{-}$auto performs better 
than the other SExtractor choices, as long as a careful match between the different catalogs is performed. We  
matched the coordinates of the sources in the 3-bands within 1$''$ to 5$''$ and after visually checking each identification we chose 3$''$ as our
best matched catalog with FUV, NUV and R-detections. The cross-match in the three bands resulted not only in the four 
most luminous members of HCG 100 (Table 1) but also  16 other objects, within the 10$'$$\times$10$'$ field (Table 2).

Nine of these sources are within the H{\sc i} tail: \# 3 is the most distant object from the group (381 arcsec or 137.2 kpc from HCG100a) and is located in the 
southern tip of the H{\sc i} tail, \#4 is far from all of the brightest members of the group and located in the densest H{\sc i} region, \#5 and
\#6 are close to HCG100c, \#8 is isolated and 150 arcsec (53.9 kpc) from HCG100a, \#9 is close to HCG100a, 
\#13 is a small galaxy not originally classified as a member of the HCG100 
and located in the optical tidal tail of HCG100b, \#14 and \#15 are also close to the tidal tail. 
The other six objects have no H{\sc i} detection but are still within the chosen field.

SE's magnitudes in FUV, NUV and R (mag$_{-}$auto) were corrected for foreground Galactic extinction using E(B-V)=0.081 and A$_{R}$=E(B-V)$\times$2.634 
(Schlegel et al. 1998), A$_{FUV}$=E(B-V)$\times$8.29 and A$_{NUV}$=E(B-V)$\times$8.18 (Seibert et al. 2005).  

In Table 3 we list the H{\sc i} mass per beam in the vicinity of nine of the intragroup objects detected, the other 
7 objects are below the detection limit ($<1.7 \times 10^{8}$). However, it is important to note that, due to the low resolution of the
data, the HI masses of the objects near large galaxies are contaminated
by the HI masses of the latter. Only objects \#3, \#4 and \#8 are far
away from bright members of the group and this contamination could
be avoided.  Objects \#13, \#14, \#15 besides being close to
H100b are within the same beam, therefore the HI masses listed in Table 3 
are not the individual masses of each object but the total mass within one 
beam centered in that region.

As seen in Fig.\ref{fuvhi_10arcmin} most of the intragroup objects are located in the outskirts of the
H{\sc i} contours where column densities range from 7.5 $\times 10^{19}$ to 5$\times 10^{20}$ cm$^{-2}$, 
except for object \#4 which is located in a peak of H{\sc i} where NH{\sc i} is 1.2$\times 10^{21}$ cm$^{-2}$. 
Therefore, H{\sc i} clouds to which many of them are associated have column densities about one order
of magnitude lower than the N(HI)$\sim$10$^{21}$ cm$^{-2}$, value thought to be required for triggering star formation (e.g. Skillman et al. 1988). 
Due to the low resolution of the HI data, all values are lower limits to the true HI column density.

\begin{deluxetable}{cccccccc}
\tablecaption{HCG100 Properties}
\tablewidth{0pt}
\tablehead{
\colhead{ID} &
\colhead{Morphology$^{\rm a}$} &
\colhead{Velocity$^{\rm b}$} &
\colhead{FUV$^{\rm c}$} &
\colhead{NUV} &
\colhead{R$^{\rm d}$} &
\colhead{FUV$_{\rm corr}$$^{\rm e}$} &
\colhead{NUV$_{\rm corr}$} 
}
\startdata
HCG100 a &  Sb & 5323 &17.30 & 16.78 & 12.5 & 16.12 & 16.63\\
HCG100 b &  Sm & 5163 &17.49 & 17.10 & 14.1 & 16.44 & 16.82\\
HCG100 c &  SBc& 5418 &18.46 & 17.91 & 14.7 & 17.25 & 17.79\\
HCG100 d &  Scd&      &19.08 & 18.65 & 15.5 & 17.99 & 18.41\\
\enddata
\tablenotetext{a}{Morphology from Plana et al. 2003}
\tablenotetext{b}{Systemic Velocity in kms$^{-1}$ from Plana et al. 2003}
\tablenotetext{c}{FUV and NUV magnitudes were obtained using IRAF task ellipse.}
\tablenotetext{d}{R-band magnitudes are from Rubin et al. 1991.}
\tablenotetext{e}{Extinction corrections using Seibert et al. (2005) for FUV and NUV.}
\end{deluxetable}

\begin{deluxetable}{ccccrrrr}
\tabletypesize{\scriptsize}
\tablecaption{FUV sources within HCG100 $10' \times 10'$ field}
\tablewidth{0pt}
\tablehead{
\colhead{ID} &
\colhead{RA} &
\colhead{Dec} &
\colhead{R$^{\rm a}$} &
\colhead{NUV} &
\colhead{FUV} &
\colhead{FUV-R} &
\colhead{FUV-NUV} 
}
\startdata
1 &    0.2410&  13.0647& 18.80 $\pm$  0.01&    20.56 $\pm$   0.04&  20.44  $\pm$   0.06&   1.64 $\pm$ 0.06 & -0.12 $\pm$ 0.07\\
2 &    0.2526&  13.1161& 19.54 $\pm$  0.01&    21.61 $\pm$   0.08&  21.48  $\pm$   0.10&   1.94 $\pm$ 0.10 & -0.13 $\pm$ 0.13\\
3 &    0.2722&  13.0236& 19.09 $\pm$  0.01&    21.61 $\pm$   0.08&  21.21  $\pm$   0.09&   2.12 $\pm$ 0.09 & -0.40 $\pm$ 0.12\\
4 &    0.2929&  13.0859& 20.25 $\pm$  0.04&    21.26 $\pm$   0.07&  21.09  $\pm$   0.09&   0.84 $\pm$ 0.10 & -0.17 $\pm$ 0.12\\
5 &    0.2933&  13.1377& 19.42 $\pm$  0.01&    22.43 $\pm$   0.15&  22.79  $\pm$   0.22&   3.37 $\pm$ 0.22 &  0.36 $\pm$ 0.27\\
6 &    0.3076&  13.1630& 17.48 $\pm$  0.00&    20.19 $\pm$   0.03&  20.20  $\pm$   0.05&   2.72 $\pm$ 0.05 &  0.01 $\pm$ 0.06\\
7 &    0.3176&  13.0290& 18.43 $\pm$  0.00&    22.85 $\pm$   0.11&  21.70  $\pm$   0.12&   3.27 $\pm$ 0.12 & -1.15 $\pm$ 0.16\\
8 &    0.3188&  13.0724& 20.39 $\pm$  0.01&    22.49 $\pm$   0.13&  21.18  $\pm$   0.11&   0.79 $\pm$ 0.11 & -1.31 $\pm$ 0.17\\
9 &    0.3429&  13.1208& 20.35 $\pm$  0.01&    22.24 $\pm$   0.11&  22.10  $\pm$   0.16&   1.75 $\pm$ 0.16 & -0.15 $\pm$ 0.19\\
10&    0.3535&  13.0704& 19.75 $\pm$  0.01&    21.92 $\pm$   0.09&  21.66  $\pm$   0.11&   1.91 $\pm$ 0.11 & -0.26 $\pm$ 0.14\\
11&    0.3627&  13.0532& 19.42 $\pm$  0.01&    21.63 $\pm$   0.07&  21.62  $\pm$   0.10&   2.20 $\pm$ 0.10 & -0.01 $\pm$ 0.12\\
12&    0.3704&  13.1375& 18.88 $\pm$  0.00&    22.15 $\pm$   0.09&  21.98  $\pm$   0.14&   3.10 $\pm$ 0.14 & -0.18 $\pm$ 0.17\\
13$^{\rm b}$&	 0.3735&  13.0986& 17.69 $\pm$  0.00&	 19.63 $\pm$   0.02&  19.59  $\pm$ 0.04&   1.90 $\pm$ 0.04 & -0.04 $\pm$ 0.04\\
13A&   0.3738&  13.0987& 17.83 $\pm$  0.01&    19.78 $\pm$   0.05&  19.81  $\pm$   0.17&   1.98 $\pm$ 0.17 &  0.03 $\pm$ 0.18\\
13B&   0.3719&  13.0984& 17.77 $\pm$  0.01&    19.62 $\pm$   0.04&  19.83  $\pm$   0.04&   2.06 $\pm$ 0.04 &  0.22 $\pm$ 0.06\\
14&    0.3773&  13.1041& 20.42 $\pm$  0.01&    21.63 $\pm$   0.08&  21.10  $\pm$   0.09&   0.69 $\pm$ 0.09 & -0.53 $\pm$ 0.12\\
15&    0.3796&  13.0950& 19.52 $\pm$  0.01&    22.04 $\pm$   0.12&  21.79  $\pm$   0.14&   2.27 $\pm$ 0.14 & -0.25 $\pm$ 0.19\\
16&    0.3816&  13.0817& 18.89 $\pm$  0.00&    21.55 $\pm$   0.06&  21.30  $\pm$   0.11&   2.41 $\pm$ 0.11 & -0.24 $\pm$ 0.12\\
\enddata
\tablenotetext{a}{Magnitudes (AB) in all bands were obtained with SExtractor Mag$_{-}$auto. Galactic extinction corrections were done 
using Schlegel et al. (1998) for R and Seibert et al. (2005) for FUV and NUV.}
\tablenotetext{b}{Object 13 was separated into two objects, A and B, using IRAF polyphot task.}
\end{deluxetable}

\begin{deluxetable}{cccccccccc}
\tabletypesize{\scriptsize}
\tablecaption{Derived Properties}
\tablewidth{0pt}
\tablehead{
\colhead{ID} &
\colhead{L$_{\rm FUV}$ (erg/s)$^{\rm a}$} &
\colhead{L$_{\rm NUV}$ (erg/s)} &
\colhead{age$^{\rm b}$} &
\colhead{SFR$_{\rm FUV}$ $^{\rm c}$} &
\colhead{SFR$_{\rm NUV}$} &
\colhead{log(M$_{\rm *}$)$^{\rm d}$} &
\colhead{log(M$_{\rm HI}$)$^{\rm e}$} &
\colhead{Sep$^{\rm f}$} &
\colhead{log I$_{1530}$$^{\rm g}$}
}
\startdata
1  &  5.90E+040  & 5.70E+040  & 3.9	 & 0.005&  0.007 & 5.0 &   & 131.2	  & 6.02 \\
2  &  2.27E+040  & 2.16E+040  & 3.8	 & 0.002&  0.003 & 4.6 &   & 102.2	  & 5.91 \\
3  &  2.91E+040  & 2.17E+040  & $<$1	 & 0.002&  0.003 & 4.7 & 9.6 & 137.2   & 5.90 \\
4  &  3.25E+040  & 2.98E+040  & 3.3	 & 0.003&  0.004 & 4.7 & 10.4 & 60.7    & 5.35 \\
5  &  6.79E+039  & 1.02E+040  & 194.1	 & 0.001&  0.001 & 6.5 & 9.9$\dagger$ & 61.2    & 4.99 \\
6  &  7.41E+040  & 8.01E+040  & 26.9	 & 0.006&  0.010 & 6.3 & 9.2$\dagger$ & 74.7    & 6.31 \\
7  &  1.84E+040  & 6.90E+039  & $<$1	 & 0.001&  0.001 & 4.5 &     & 108.5	 & 5.51 \\
8  &  2.98E+040  & 9.62E+039  & $<$1	 & 0.002&  0.001 & 4.7 & 10.0 & 53.9    & 5.31 \\
9  &  1.28E+040  & 1.21E+040  & 3.5	 & 0.001&  0.001 & 4.6 & 9.8$\dagger$ & 17.5    & 5.80 \\
10 &  1.92E+040  & 1.63E+040  & 2.9	 & 0.002&  0.002 & 4.4 &     & 58.8	  & 5.97 \\
11 &  2.00E+040  & 2.13E+040  & 21.0	 & 0.002&  0.003 & 5.6 &     & 83.7	  & 5.85 \\
12 &  1.43E+040  & 1.31E+040  & 3.2	 & 0.001&  0.002 & 4.3 &     & 58.0	  & 5.95 \\
13 &  1.29E+041  & 1.35E+041  & 13.7	 & 0.010&  0.016 & 6.2 & 9.5$\dagger$$\dagger$ & 53.3    & 6.67 \\
13 A& 1.03E+041  & 1.36E+041  & 118.7	 & 0.008&  0.017 &     &  &  & 6.51 \\
13 B& 1.06E+041  & 1.17E+041  & 33.8	 & 0.008&  0.014 &     &  &  & 6.50 \\
14 &  3.22E+040  & 2.13E+040  & $<$1	 & 0.003&  0.003 & 4.8 & 9.5$\dagger$$\dagger$ & 56.3	& 6.30 \\
15 &  1.70E+040  & 1.46E+040  & 2.9	 & 0.001&  0.002 & 4.3 & 9.5$\dagger$$\dagger$ & 62.1    & 5.07 \\
16 &  2.67E+040  & 2.29E+040  & 2.9	 & 0.002&  0.003 & 4.5 &     & 72.0	  & 5.26 \\
\enddata
\tablenotetext{a}{FUV and NUV luminosities are in erg/s, divide by the FUV and NUV bandwidths (269\AA\ and 616\AA, respectively) to obtain in L erg/s/\AA.}
\tablenotetext{b}{Age (Myr) from FUV-NUV using Thilker et al. (2007) assuming a Milky Way internal extinction (E(B-V)=0.2).}
\tablenotetext{c}{SFR (M$_{\odot}$/yr) from Iglesias-P\'aramo et al. (2006) using FUV and NUV without correcting for internal extinction.}
\tablenotetext{d}{Stellar mass (M$_{\odot}$) obtained from Starburst99 using ages (column 4) and L$_{\rm 1530}$ (erg/s/\AA).}
\tablenotetext{e}{MHI (M$_{\odot}$) was calculated using 2.36 $\times$ 10$^{5}$ F$_{\rm HI}$ D$^{2}$, where D is in Mpc and F$_{\rm HI}$ in Jy Km/s. F$_{\rm HI}$ was measured within one 
beam size 61.0$''$ $\times$ 55.23$''$ and reflects the HI mass in the vicinity of each object.}
\tablenotetext{f}{Distance (kpc) between each object and HCG100a.}
\tablenotetext{g}{FUV surface brightness (L$_{\odot}$ kpc$^{-2}$) is defined following Hoopes at al. (2007).}
\tablenotetext{\dagger}{HI masses of Objects \#5, \#6, and \#9 should be taken with caution due to contamination by large galaxies in the vicinity.}
\tablenotetext{\dagger\dagger}{HI masses of Objects \#13, \#14, and \#15 are not individual masses of each object but the total mass within one 
beam centered in that region. Contamination from H100b mass is also possible.}
\end{deluxetable}





\section{Analysis}

We used the broad-band colors generated from stellar populations models by Thilker et al. (2007) 
to estimate the ages of the intragroup sources. Thilker et al. models are optimized for 
GALEX filters transmission curves and have been extensively tested using GALEX data (Bianchi et al. 2007). 

\begin{figure}
\plotone{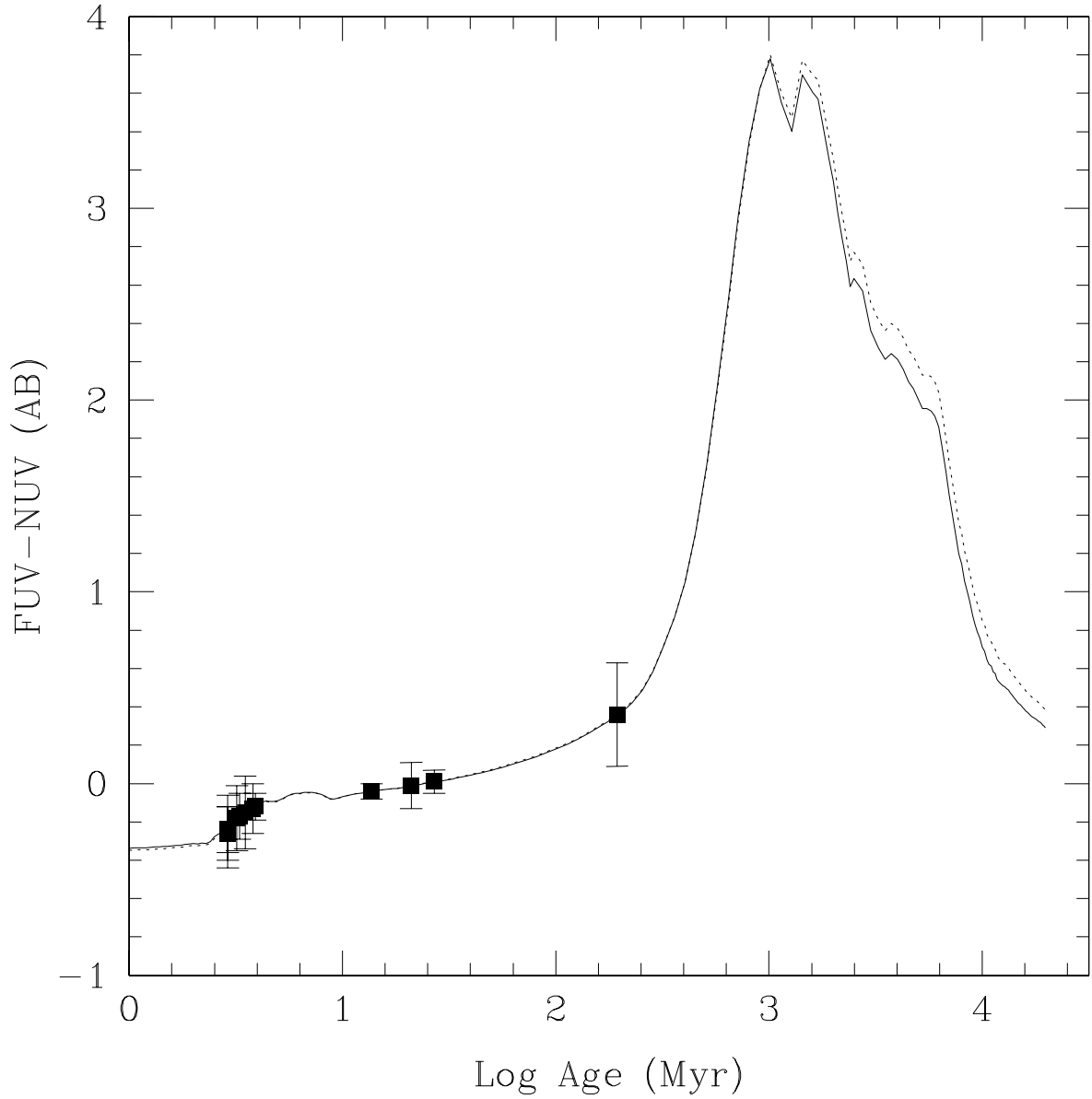}
\caption{GALEX FUV-NUV versus age from Thilker et al. models (2007) are shown as solid line (no extinction
correction) and dotted line (internal extinction of the Milky Way E(B-V)=0.2). 4 objects are not included, they have ages $<$ 1~Myr old.
\label{colorsfuvnuvthilker}}
\end{figure}

\begin{figure}
\plotone{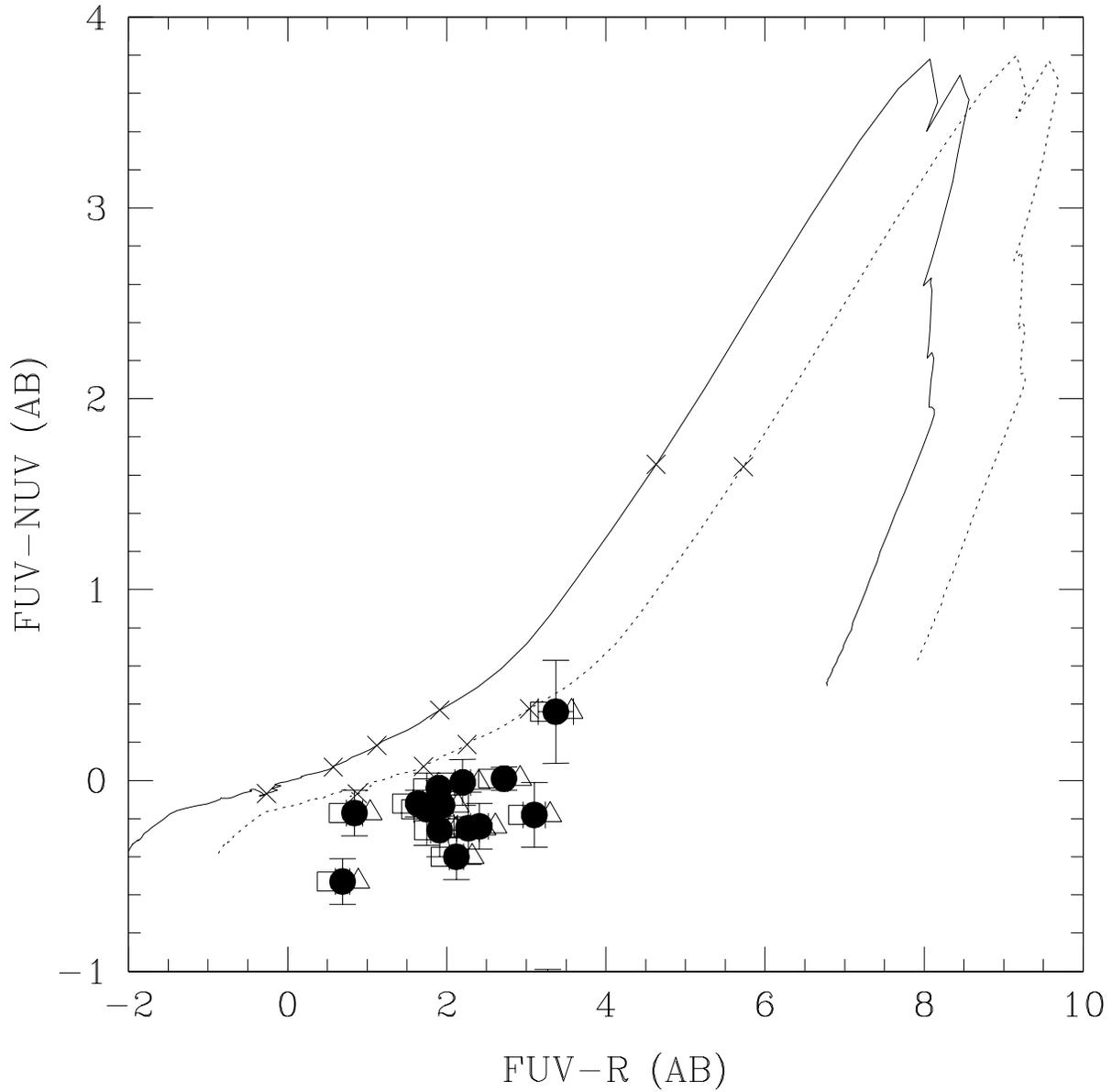}
\caption{GALEX FUV-NUV versus FUV-R of the intragroup objects. Squares and triangles are FUV--R colors assuming an error of 0.2 magnitudes in the R-magnitude. 
Models from Thilker et al. 2007 are shown as solid line (no extinction correction) and dotted line (internal extinction of the Milky Way E(B-V)=0.2). Crosses mark ages 10, 50, 100, 200, and 500 Myr, from the 
bottom left to the top right. 
\label{colorsfuvnuvfuvr}}
\end{figure}

In Fig.\ref{colorsfuvnuvthilker} we plot FUV--NUV colors of our sample and 
FUV--NUV color predictions  as a function of age assuming a Milky Way internal extinction, (E(B-V)=0.2) and
without any extinction correction (Thilker et al. 2007). It is clear that all intragroup 
objects are younger than $\sim$ 200 Myr. Four objects are not included in the logarithmic scale of Fig.\ref{colorsfuvnuvthilker}, 
they have ages $<$ 1~Myr old and are probably more affected by extinction since the first 2 Myrs after the stellar birth are strongly affected by 
dust obscuration. In Table 3
we list their ages, assuming a Milky Way internal extinction (E(B-V)=0.2) for all objects. However, for many of them the internal extinction is higher
than the Milky Way value, as seen in the color-color plot shown in Fig.\ref{colorsfuvnuvfuvr}. We have not attempted to estimate the 
intrinsic extinction individually since the three colors we have are either too close to each other (FUV and NUV) or too far apart (NUV and R). 
In any case, the ages of all objects are $<$ 200 Myr and will not reach older ages even if assuming extinction values such as A$_{\rm FUV}$=2.03,  
A$_{\rm NUV}$=1.46 used to age-dating Sloan Digital Sky Survey (SDSS) galaxies detected with GALEX  (Salim et al. 2005). The
UV-detection of these objects suggests that the energetic O-star winds have blown out the natal cocoon veiling most of the massive stars (Leitherer
2005). In Fig.\ref{colorsfuvnuvfuvr} we also plot FUV-R colors assuming an error of 0.2 magnitudes in the R-band magnitudes (triangles and squares)
confirming that an object 0.2 magnitudes brighter or fainter in R have slightly higher or lower predicted ages if we had used FUV-R as an 
age indicator. Since we used FUV-NUV as an age indicator our results are not affected by errors in the photometry of the R-band image.

As discussed in Kennicutt (1998) the broadband optical luminosity by itself is a poor tracer of the star formation rate (SFR) 
and the UV wavelengths where the integrated spectrum is dominated by young stars gives a better estimate of the SFR. In 
Table 3 we list the SFRs which were derived following Iglesias-P\'aramo et al. (2006)\footnote{SFRs are derived using 
Starburst99 models (solar metallicity, Salpeter IMF from 0.1 to 100 M$_{\odot}$) adapted to the GALEX bands.} without applying internal  
extinction correction to the UV flux. 
The SFR using FUV and NUV are similar except for objects \#6 and \#13. This could be due to the fact that 
the latter object has double nuclei and is probably  a merger of two small objects, whereas the former object is the largest 
object in the sample and probably a dwarf galaxy. Therefore, the instantaneous star-formation law used to derive the SFR might not be suited 
to their stellar populations. If they have had several bursts of star-formation in the recent past, the FUV gives a more
robust indicator of SFR than NUV since it is not contaminated by previous star formation. As shown in Salim et al. (2005) the NUV-R is a good tracer of the 
star formation
history of a galaxy, i.e. it is proportional to the ratio between the current SFR over the past-average SFR. If we assume Salim et al. average value
of extinction (A$_{\rm NUV}$=1.46), all objects in our sample have NUV-R typical of galaxies undergoing a burst.

We have also calculated SFR$_{\rm FUV}$ using Calzetti (2001) values of extinction for starburst galaxies. The average SFR$_{\rm FUV}$ of the intragroup 
objects is as low as 0.01~M$_{\odot}$ yr$^{-1}$, and as high as 0.4~M$_{\odot}$ yr$^{-1}$, if we use the A$_{1500}$ = 1.6 (Calzetti's UV-selected starbursts) 
and A$_{1500}$ = 4.1 (Calzetti's average dust-rich starbursts), respectively. Therefore, the SFR of the intragroup objects is modest even if a high 
extinction correction is applied. 

We used the ages obtained from Thilker et al. and the FUV luminosity to estimate the stellar masses with Starburst99 models (Leitherer et al. 1999).
The stellar masses were obtained from Starburst99's monochromatic luminosity, L$_{1530}$, for an instantaneous burst, Salpeter IMF, solar metallicity, and 
standard input parameters are listed in Table 3. We chose an instantaneous burst instead of continuous star-formation, since we are estimating the 
mass from the FUV luminosity which is dominated by the young and massive stars. Therefore, low mass stars were not taken into consideration in the stellar mass 
calculation.

\subsection{Comparison with Intergalactic HII Regions and TDGs}

Tidal dwarf galaxies (TDGs) are found in the tip of the optical tails as well as along tidal features 
(e.g. Duc et al. 1997, Weilbacher et al. 2000, Knierman, et al. 2003, Bornaud et al. 2004). 
More recently, they have been identified in ongoing galaxy mergers where the FUV emission is seen to be associated with the H{\sc i} tidal 
tails (Hibbard et al. 2005), extending 
large distances beyond the optical extent (Neff et al. 2005). One of our objects, 
\#4, is similar to Neff et al.'s TDGs. It is located in the H{\sc i} tail and where the H{\sc i} column density peaks. 
It has an H{\sc i} mass 10$^{10.4}$ M$_{\odot}$/beam in the vicinity  and 
age 3.3~Myr. What is interesting for this object and also object \#3, is their location with respect to the 
 group members; 
object \#4 is 61 kpc away from HCG100a and object \#3 is at the very tip of the HI tail, 137 kpc away from HCG100a. 

The TDG-candidate found in HCG100 (object \#4) described above and other TDGs found in other systems with H{\sc i} tails are located far away from the
group members and have young ages. For instance, the triplet formed by 
NGC7771,  NGC7770 and NGC7769 observed with GALEX has a 3~Myr old TDG candidate within the 100~kpc H{\sc i} tail (Neff et al 2005). 
Another compact group, HCG31 has a TDG 40 kpc away from the parent galaxy (Mendes de Oliveira et al. 2006) and is also located in a peak of H{\sc i} in the gas tail .

The tidal material in interacting systems is expected to remain bound to the system. Some of it will fall back onto the remnant and some
will stay detached ($\sim$20\% of NGC7252 tidal material will stay detached, according to Hibbard \& Mihos 1995) for as long as a Hubble 
time. According to Verdes-Montenegro et al.'s (2005) proposed evolutionary scenario for compact groups, HCG100 is in phase 2, when multiple tidal tails form
 and a large amount of atomic gas is found in the intragroup medium. In phase 3 of the evolutionary path, galaxies will lose H{\sc i} 
 gas and an H{\sc i} deficient group will be produced (Sulentic et al. 2001, Williams et al. 2002). 
Based on our results we suggest that at least a fraction of the tidal gas is used to form intragroup objects such as the ones within the H{\sc i} tail.

Moreover, considering the distance between the star forming regions and the group members and the young ages of the stellar population, we are witnessing the in ``situ" 
star-formation within the H{\sc i} tidal tail and can rule out the possibility of formation in
the outskirts of the group galaxies and later ejection to the intragroup medium.

\begin{figure}
\plotone{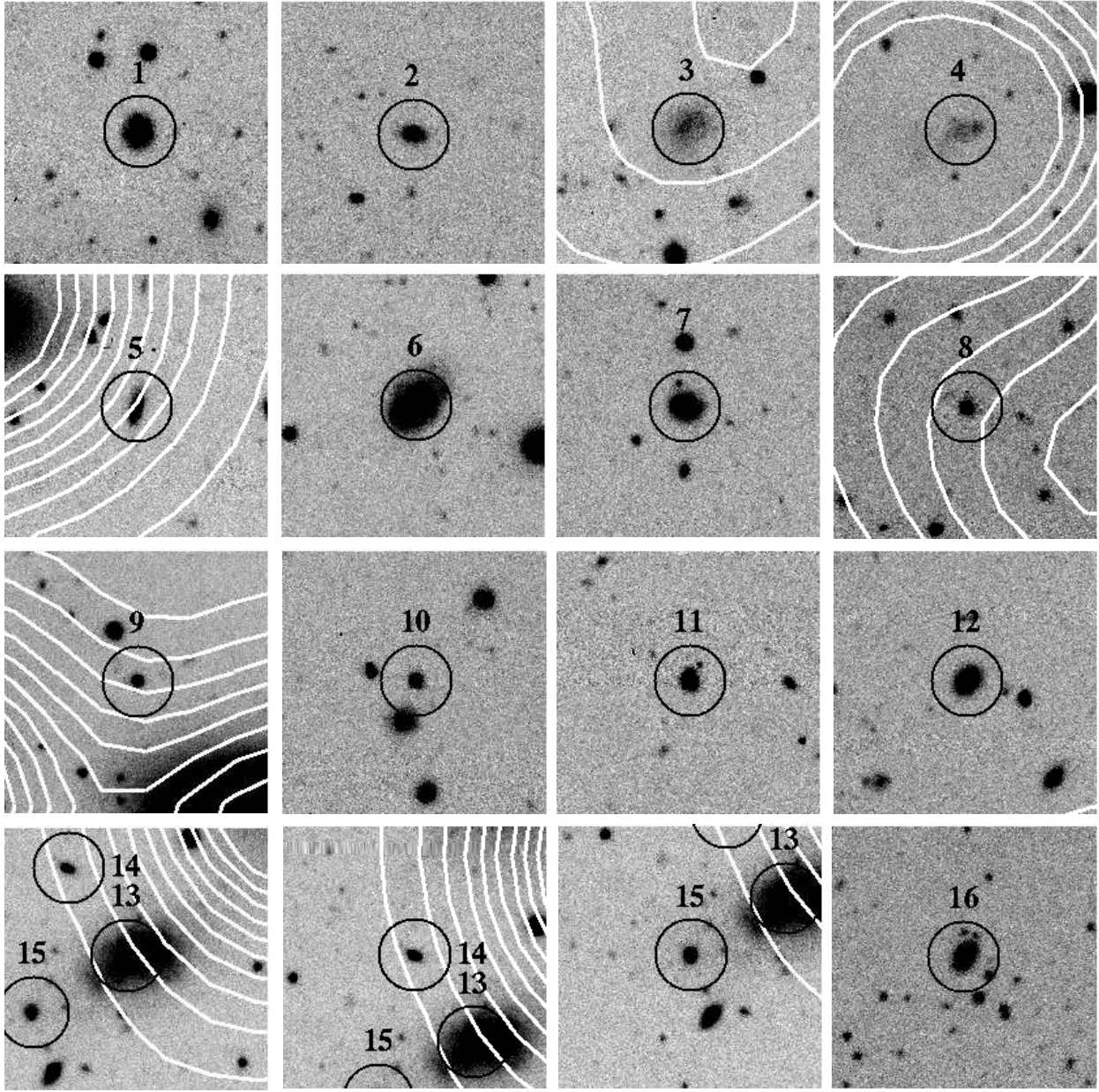}
\caption{R images of all intragroup objects, circle has radius=8$''$, each cutout is 1$'$$\times$1$'$. NH{\sc i} contours (white) are the same as in Fig.\ref{fuvhi_10arcmin}.
\label{all_obj}}
\end{figure}

As shown in the R-band images (Fig.\ref{all_obj}), objects \#3 and \#4 resemble low surface brightness galaxies, 
whereas the others are either small round objects (\#8, \#9, \#10, and \#15), but larger than the images' PSF, 
or have morphologies similar to objects found in the tail region
of Stephan's Quintet (Mendes de Oliveira et al. 2004). We cannot rule out the possibility that these objects 
are not related to the group and only with spectroscopy we will be
able to tell whether they are group members or not. However, if found to be bound to HCG100, as the intergalactic H{\sc ii} regions 
in Stephan's Quintet are, the intragroup medium of HCG100 is actively forming stars as far as 137 kpc away from the galaxies. The SFRs are modest,  
0.001--0.02~M$_{\odot}$ yr$^{-1}$, but similar to IH{\sc ii} regions found by Ryan-Weber et al. (2004) at distances up to 33 kpc from NGC~1533 and
by Sakai et al. (2002) at distances of 155 kpc in the A1367 cluster. 

The stellar masses of our sources (Table 3) range between 10$^{4.3}$--10$^{6.5}$ M$_{\odot}$. The large spread in masses can be interpreted as a
sign of evolution or different population of objects. In the first scenario, some of the objects are still in the process of formation and have not converted 
all the gas supply into stars as Duc \& Mirabel (1997) suggested for TDGs. In the second scenario, we are detecting two populations of
objects: TDGs and IH{\sc ii} regions. The low--mass objects are the IH{\sc ii} regions and high--mass objects are TDGs. 
One caveat to keep in mind is the fact that the stellar masses can be higher than what we obtained, since the intrinsic extinction for some objects 
is probably higher than the Milky Way value used to estimate their age. Higher extinction correction would make them more UV luminous 
and therefore more massive. 

\subsection{Comparison with high-z galaxies}

Since the intragroup objects were UV-selected, we have compared them with a class of objects which is also
selected by their UV emission but are at high redshifts, the Lyman Break Galaxies (LBGs, Steidel et al. 1995). LBGs are star-forming 
galaxies at $z=2-4$, they are relatively small (r$_{h}$=1-3 kpc), 
have relatively low mass (10$^{9.5-11}$M$_\odot$) and low extinction. LBGs are thought to be the building blocks of larger 
systems proposed by the hierarchical clustering scenario of galaxy formation and evolution (Kauffmann et al. 1993).  The fate 
of intragroup objects might be similar, since at least some are likely to be accreted by the larger galaxies in the group. HCG100a, for instance, 
may have already suffered a major accretion event, given its very peculiar velocity field (Plana et al. 2003). 

Recently, Hoopes et al. (2007) identified a sample of the local equivalents of LBGs using images from the UV satellite GALEX and 
spectroscopy from the SDSS. Among the UV-luminous galaxy (UVLGs) population, they 
found a subset of supercompact galaxies that are similar to LBGs. As shown in Table 4, the intragroup objects in HCG100 are not as UV-luminous 
as UVLGs and LBGs and also form stars in a much lower rate than those. This might be an evolutionary effect suggesting that objects in the 
intragroup medium are less efficient in forming stars than LBGs and UVLGs.

\section{Conclusions}

We have analyzed the UV and optical light in combination with the H{\sc i} gas within the compact group of galaxies HCG100 and
identified 16 star-forming regions in the intragroup region. 
The young age ($<200$~Myr) of these objects and the proximity to the 
tidal tail connects 
the OB stars formation time scale ($\sim$10$^{8}$ yr) with the dynamic time scale of the tidal features. 
Moreover, the H{\sc i} clouds to which many of them are associated have column densities about one order
of magnitude lower than the N(HI)$\sim$10$^{21}$ cm$^{-2}$
thought to be required for triggering star formation. So,
in these cases, we have a strong indication that the H{\sc i} clouds
must have suffered recent collisions which could have then triggered
the star formation process. Based on their ages, stellar masses and H{\sc i} masses in their vicinities, we conclude that some of these objects are tidal dwarf 
galaxies with ongoing star formation and some are intergalactic HII regions or conglomeration of stellar clusters.

Although we have an estimate of the amount of neutral gas in these objects, the main ingredient in the star-formation process, the 
molecular gas, is unknown for these objects. From CO observations of tidal dwarf galaxies (TDGs), Braine et al. (2001) provided strong 
evidence that TDGs are self-gravitating entities with large amounts of atomic gas which will transform into molecular gas and subsequently 
form stars. CO observations together with spectroscopy of these objects will be of great value in understanding their nature. 
For instance, we cannot exclude the possibility that some of these objects are not associated with HCG100, 
but at least those falling within HI density peaks are likely to be associated with the group. Moreover, 
we will be able to estimate metallicity of these newly formed regions and evaluate the enrichment level of the processed
gas in the intragroup medium.

\begin{deluxetable}{lcccc}
\tablecaption{Comparison of Galaxy Properties}
\tablewidth{0pt}
\tablehead{
\colhead{Parameter$^{\rm a}$} &
\colhead{Large UVLGs} &
\colhead{Compact UVLGs} &
\colhead{LBGs} &
\colhead{HCG100 IGs} 
}
\startdata
log L$_{1530}$ (L$_{\odot}$)                     & 10.3 to 11.2  & 10.3 to 11.0  & 10.3 to 10.9 &  8 to 9 \\
log I$_{1530}$ (L$_{\odot}$ kpc$^{-2}$) $^{\rm b}$& 6.0 to 8.0    & 8.0 to 10.3   & 9.0 to 10.3 &  5.6 to 8.3 \\
log SFR (M$_{\odot}$ yr$^{-1}$) $^{\rm c}$       & 0 to 1.5      &  0.2 to 2.0   & 0.5 to 2.0   & -2 to -0.4   \\
FUV-R   $^{\rm d}$                               & 1.0 to 3.5    &  0.2 to 2.8   & 0.2 to 1.7   &  0.7 to 3.4 $^{\rm f}$ \\
\enddata
\tablenotetext{a}{Parameters for UVLGS and LBGs are from Hoopes et al. (2007). The lowest value for HCG100 intragroup objects was 
calculated using average values A$_{1500}$=1.6 and the highest value using A$_{1500}$=4.1 from Calzetti (2001).}
\tablenotetext{b}{I$_{1530}$ is the FUV surface brightness as described in Hoopes et al. (2007), except for HCG100 intragroup objects where we used 
R as the FUV petrosian radius.}
\tablenotetext{c}{SFR for the HCG intragroup objects using FUV equation from Iglesias-P\'aramo et al. (2006).}
\tablenotetext{d}{r magnitudes are from SDSS and corrected for extinction, except for HCG100 which is the same as in Table 1 (Cousin R magnitude) and is not 
corrected for extinction.}
\tablenotetext{e}{FUV-R shown is not corrected for internal extinction. Using Calzetti (2001) A$_{1500}$=1.6 and  A$_{r}$=0.48 FUV-R=-0.3 to 2.2.}
\end{deluxetable}

\clearpage

\clearpage

\acknowledgments

We are grateful to L. Verdes-Montenegro for making the reduced VLA data available, to GALEX and NOAO staff, and to the anonymous referee for helping improving the
paper. DFdM was funded by Nasa Research Grants NNG06GG45G and NNG06GG59G, ST-F was a CNPq MSc student fellow at IAG/USP and currently is a CAPES PhD 
student fellow at IAG/USP.
This research has made use of the NASA/IPAC Extragalactic Database (NED) which is operated by the Jet Propulsion Laboratory, 
California Institute of Technology, under contract with the National Aeronautics and Space Administration.


{\it Facilities:} \facility{Galex}, \facility{CTIO}.



\end{document}